\documentclass{ifacconf}
\makeatletter\let\proof\@undefined\let\endproof\@undefined\makeatother
\usepackage{amsmath}

\usepackage{amssymb}
\usepackage{xcolor}
\usepackage{enumitem}
\usepackage{comment}
\usepackage{graphicx} 
\usepackage{fixltx2e}
\usepackage{caption}
\usepackage{subcaption}
\usepackage{natbib}
\usepackage{breqn}
\usepackage{stfloats}

\usepackage{amsfonts}		% allows further math symbols
\let\proof\relax 
\let\endproof\relax
\usepackage{mathtools}		% allows further math commands
\usepackage{cases}     	% brace equations with multi-sublabels
\usepackage{bm}

\newcommand{\B}[1]{\if#1\relax\bm{#1}\else\mathbf{#1}\fi} % bold text
\newcommand{\R}{\mathbb{R}}
\newcommand{\sL}{\mathcal{L}}
\newcommand{\sN}{\mathcal{N}}
\newcommand{\sKL}{\mathcal{KL}}
\newcommand{\sK}{\mathcal{K}}
\newcommand{\abs}[1]{\vert #1 \vert}
\newcommand{\norm}[1]{\Vert #1 \Vert}

\newcommand{\T}{^{\mathsf{T}}}

\newtheorem{definition}{Definition}

\newtheorem{lemma}{Lemma}

\newtheorem{remark}{Remark}
\newtheorem{proposition}{Proposition}
\makeatletter
\DeclareRobustCommand{\qed}{%
  \ifmmode % if math mode, assume display: omit penalty etc.
  \else \leavevmode\unskip\penalty9999 \hbox{}\nobreak\hfill
  \fi
  \quad\hbox{\qedsymbol}}
\newcommand{\openbox}{\leavevmode
  \hbox to.77778em{%
  \hfil\vrule
  \vbox to.675em{\hrule width.6em\vfil\hrule}%
  \vrule\hfil}}
\newcommand{\qedsymbol}{\openbox}
\newenvironment{proof}[1][\proofname]{\par
  \normalfont
  \topsep6\p@\@plus6\p@ \trivlist
  \item[\hskip\labelsep\itshape
    #1.]\ignorespaces
}{%
  \qed\endtrivlist
}
\newcommand{\proofname}{Proof}
\makeatother

\begin{document}
\begin{frontmatter}

\title{On the design of scalable networks rejecting first order disturbances}%\thanksref{footnoteinfo}} 
% Title, preferably not more than 10 words.

\thanks[footnoteinfo]{The work of Shihao Xie was supported by the Science Foundation Ireland (SFI) under Grant $16/IA/4610$.}

\author[First]{Shihao Xie} 
\author[Second]{Giovanni Russo} 

\address[First]{School of Electrical and Electronic Engineering, University College Dublin, Ireland (e-mail: shihao.xie1@ucdconnect.ie)}
\address[Second]{Department of Information and Electrical Engineering and Applied Mathematics, University of Salerno, Italy (e-mail: giovarusso@unisa.it)}

\begin{abstract}    % Abstract of not more than 250 words.
This paper is concerned with the problem of designing distributed control protocols for network systems affected by delays and disturbances consisting of a first-order polynomial component and a residual signal. Specifically, we propose the use of a multiplex architecture to design distributed control protocols to reject polynomial disturbances up to ramps and  guarantee a scalability property that prohibits the amplification of  residual disturbances. For this architecture, we give a sufficient condition on the control protocols to guarantee scalability and ramps rejection. The effectiveness of the result, which can be used to study networks of nonlinearly coupled nonlinear agents, is illustrated via a robot formation control problem.
\end{abstract}

\end{frontmatter}

\section{Introduction}
%A considerable amount of research effort has been devoted to the study of agents achieving desirable collaborative objectives, including consensus, synchronization, formation control etc.\\
%Network systems have significantly increased both their size and complexity over the years. Examples of large-scale networks include e.g. smart grids, traffic networks, biological systems and neural networks. \\, 1101721
%Network system is a group of coordinated agents aiming at achieving desirable collaborative objectives, including consensus, synchronization, formation control etc. Examples of networks include e.g. smart grids, traffic networks, biological systems and neural networks. \\
Over the last few years, network systems have considerably evolved, increasing  their size and complexity of their topology. The study of coordinated behaviours, such as consensus and synchronization, has therefore attracted much research attention \citep{6891349, 5530690}. In this context, a key challenge is the design of protocols that do not only guarantee {\em stability} (i.e. the fulfillment of the desired, coordinated behavior) but also: (i) ensure rejection of certain classes of disturbances; (ii) guarantee that the network is {\em scalable} with respect to disturbances that are not fully rejected, i.e. disturbances that are not rejected are not amplified across the network. We use the word {\em scalability} to denote the preservation of the desired properties (to be defined formally in Section \ref{sec:control_goal}) uniformly with respect to the number of agents. Disturbances can be often modeled as the sum of a polynomial component \citep{6425973} and a residual signal, capturing components that cannot be modeled via a polynomial. Motivated by this, we: (1) propose a multiplex \citep{BURBANOLOMBANA2016310}  architecture (defined in Section \ref{sec:problem_set-up}) with the aim of simultaneously guaranteeing rejection of polynomial disturbances up to ramps and scalability   for nonlinear networks affected by delays; (2) give a sufficient condition on the control protocol to assess these properties; (3) illustrate the effectiveness of the result on a formation control problem.

\subsubsection*{Related works.}
The study of how disturbances propagate within a network is a central topic for autonomous vehicles. In particular, the key idea behind the several definitions of string stability \citep{swaroop} in the literature is that of giving upper bounds on the deviations induced by disturbances that are uniform with respect to platoon size, see e.g. \citep{KNORN20143224, 6515636, 7879221, MONTEIL2019198} for a number of recent results. These works assume delay-free inter-vehicle communications and an extension to delayed platoons can be found in e.g. \citep{6891349}. For networks with delay free interconnections, we also recall here results on mesh stability \citep{Seiler1999PreliminaryIO} for networks with linear dynamics and its extension to nonlinear networks in \citep{mesh}. Leader-to-formation stability is instead considered in \citep{1303690} and it characterizes network behavior with respect to inputs from the leader. For delay-free, leaderless networks with regular topology, scalability has been recently investigated in \citep{8370724}, where Lyapunov-based conditions were given; for networks with arbitrary topology and delays, sufficient conditions for scalability are given in \citep{9353260} leveraging non-Euclidean contraction, see e.g. \citep{LOHMILLER1998683,1618853,8561231} and \citep{7937859} where contraction analysis was first used in the context of platooning. Finally, we recall that in the context of vehicle platooning, the problem of guaranteeing string stability and simultaneously rejecting constant disturbances has been investigated in  \citep{KNORN20143224,SILVA2021109542} and this has led to the introduction of an integral action in the control protocol.

\subsubsection*{Statement of contributions.}
We tackle the problem of designing network systems that are both scalable and are also able to reject polynomial disturbances up to ramps. Our main contributions can be summarized as follows: (i) for possibly nonlinear networks affected by delays, we propose  a multiplex  architecture to guarantee both rejection of ramp disturbances and scalability (with respect to any residual disturbances) requirements. To the best of our knowledge, this is the first work to introduce the idea of leveraging multiplex architectures for disturbance rejection; (ii) the main result we present, which applies to both leader-follower and leaderless networks, is a sufficient condition guaranteeing the fulfillment of the {\em ramp-rejection} and scalability requirements. We are not aware of other results to fulfill these requirements; (iii) the result is then turned into a  design guideline and its effectiveness is illustrated on a formation control problem.

\section{Mathematical preliminaries}
Let $A$ be a $m \times m$ real matrix, we denote by $\norm{A}_p$ the matrix norm induced by the $p$-vector norm $\abs{\cdot}_p$. The matrix measure of $A$ with respect to $\abs{\cdot}_p$ is defined by $\mu_p(A)=\lim_{h\rightarrow 0^+}\frac{\norm{I+hA}_p-1}{h}$. Given a piecewise continuous signal $w_i(t)$, we let $\norm{w_i(\cdot)}_{\sL_\infty^p}:=\sup_t\abs{w_i(t)}_p$. We denote by $I_n$ ($0_n$) the $n\times n$ identity (zero) matrix and by $0_{m\times n}$ the $m\times n$ zero matrix. We let $\textbf{diag}\{a_1,\ldots,a_N\}$ be a diagonal matrix with diagonal elements $a_i$. Given a generic set $\mathcal{A}$, its cardinality is denoted as $\textbf{card}(\mathcal{A})$. We recall that a continuous function $\alpha: [0,a)\rightarrow [0,\infty)$ is said to belong to class $\sK$ if it is strictly increasing and $\alpha(0)=0$. It is said to belong to class $\sK_\infty$ if $a=\infty$ and $\alpha(r)\rightarrow \infty$ as $r \rightarrow \infty$. A continuous function $\beta: [0,a)\times [0,\infty)\rightarrow [0,\infty)$ is said to belong to class $\sKL$ if, for each fixed $s$, the mapping $\beta(r,s)$ belongs to class $\sK$ with respect to $r$ and, for each fixed $r$, the mapping $\beta(r,s)$ is decreasing with respect to $s$ and $\beta(r,s)\rightarrow 0$ as $s\rightarrow \infty$.

Our results leverage the following lemma, which can be found in \citep{9353260} and follows directly from \citep{5717887}. To state the result we let $\abs{\cdot}_S$ and $\mu_S(\cdot)$ be, respectively, any monotone norm and its induced matrix measure on $\mathbb{R}^N$. In particular, we say a norm $\abs{\cdot}_S$ is monotone if for any non-negative $N$-dimensional vector $x,y \in \R_{\ge 0}^N$, $x\le y$ implies that $\abs{x}_S \le \abs{y}_S$ where the inequality $x\le y$ is component-wise.
\begin{lemma}\label{lem:matrix norm}
Consider the vector $\eta:=[\eta_1\T,\ldots,\eta_N\T]\T$, $\eta_i \in \mathbb{R}^n$. We let $\abs{\eta}_G := \abs{\left[ \abs{\eta_1}_{G_1},\ldots,\abs{\eta_N}_{G_N}\right]}_S$, with $\abs{\cdot}_{G_i}$ being norms on $\R^n$, and denote by $\norm{\cdot}_G, \mu_G(\cdot)$ ($\norm{\cdot}_{G_i}, \mu_{G_i}(\cdot)$) the matrix norm and measure induced by $\abs{\cdot}_G$ ($\abs{\cdot}_{G_i}$). Finally, let: (1) $A:=({A_{ij}})_{i,j=1}^N \in \mathbb{R}^{nN \times nN}$, $A_{ij} \in \mathbb{R}^{n \times n}$; (2) $\hat{A}:=(\hat{A}_{ij})_{i,j=1}^N \in \mathbb{R}^{N \times {N}}$, with $\hat{A}_{ii}:=\mu_{G_i}(A_{ii})$ and $\hat{A}_{ij}:=\norm{A_{ij}}_{G_{i,j}}$, {$\norm{A_{ij}}_{G_{i,j}}:=\sup_{\abs{x}_{G_i}=1}\abs{A_{ij}x}_{G_j}$};(3) $\bar{A}:=(\bar{A}_{ij})_{i,j=1}^N \in \mathbb{R}^{N \times N}$, with $\bar{A}_{ij}:=\norm{A_{ij}}_{G_{i,j}}$. 
Then: (i) $\mu_G(A) \le \mu_S(\hat{A})$; (ii) $\norm{A}_G \le \norm{{\bar A}}_S$. 
\end{lemma}
We recall here that, if the norms $\abs{\cdot}_S$, $\abs{\cdot}_{G_{1}}, \ldots, \abs{\cdot}_{G_{N}}$ in Lemma \ref{lem:matrix norm} are $p$-norms (with the same $p$) then $\abs{\cdot}_G$  is again a $p$-norm (although defined on a larger space). The next lemma follows from Theorem $2.4$ in \citep{WEN2008169}.
\begin{lemma}\label{prop:halanay}
Let $u:[t_0-\tau_{\max},+\infty)\rightarrow\R_{\ge 0}$ , $\tau_{\max}<+\infty$ and assume that
$$
D^+u(t) \le au(t)+b \sup_{t-\tau(t) \le s \le t}u(s)+c, \ \  t\ge t_0
$$
with: (i) $\tau(t)$ being bounded and non-negative, i.e. $0{\le}\tau(t)\le\tau_{\max}$, $\forall t$; (ii) $u(t)=\abs{\varphi(t)}$, $\forall t\in [t_0-\tau_{\max},t_0]$ where $\varphi(t)$ is bounded in $[t_0-\tau_{\max},t_0]$; (iii) $a < 0$, $b \ge 0$ and $c \ge 0$. Assume that there exists some $\sigma>0$ such that $a+b \le -\sigma <0, \forall t\ge t_0$. Then:
$$
u(t) \le \sup_{t_0-\tau_{\max} \le s \le t_0}u(s)e^{-\hat \lambda (t-t_0)}+\frac{c}{\sigma},
$$
where $\hat \lambda:=\inf_{t\ge t_0}\{\lambda|\lambda(t) + a+be^{\lambda(t)\tau(t)}=0\}$ is positive.
\end{lemma}

\section{Statement of the control problem}\label{sec:problem_set-up}
%We start with formally stating the control problem considered in this paper. To this aim, we first give a description of the control architecture we propose (Section \ref{sec:architecture}) and then formalize the control goal (Section \ref{sec:control_goal}).

%\subsection{Architecture description}\label{sec:architecture}
We consider a network system of $N>1$ agents with the dynamics of the $i$-th agent given by
\begin{align}\label{equ: dynamics}
\begin{split}
    \dot{x}_i(t)&=f_i(x_i,t)+u_i(t)+d_i(t), \ \ t\ge t_0\ge 0,\\
    %y_i(t)&=g_i(x_i),
\end{split}
\end{align}
%\SX{i think we don't need the output function}
with $i=1,\dots,N$, initial conditions being $x_i(t_0)$ and where: (i) $x_i(t)\in \R^n$ is the state of the $i$-th agent; (ii) $u_i(t) \in \R^n$ is the control input; (iii) $d_i(t) \in \R^n$ is an external disturbance signal on the agent; (iv) $f_i: \R^n\times \R_{\ge 0} \rightarrow \R^n$ is the intrinsic dynamics of the agent, assumed to be smooth. % (v) $g_i:\R^n\rightarrow\R^q$ is the output function for the $i$-th agent. 
We consider disturbances of the form: 
\begin{align}\label{equ: disturbance}
    d_i(t) = w_i(t) + \bar{d}_i(t) := w_i(t)+\bar{d}_{i,0}+\bar{d}_{i,1}\cdot t,
\end{align}
where $w_i(t)$ is a piecewise continuous signal and $\bar{d}_{i,0}, \bar{d}_{i,1}$ are constant vectors. Disturbances of the form of (\ref{equ: disturbance}) can be thought of as the superposition of the ramp disturbance $\bar{d}_i(t):=\bar{d}_{i,0}+\bar{d}_{i,1}\cdot t$ and the signal $w_i(t)$. In the special case when $\bar{d}_{i,1}$ is zero, \eqref{equ: disturbance} becomes $d_i(t) = w_i(t) + \bar{d}_{i,0}$ and scalability properties with respect this disturbance have been studied in the context of vehicle platooning: in \citep{SILVA2021109542}, the term $\bar{d}_{i,0}$ models the constant disturbance to the acceleration when the vehicle hits a slope and the residual term $w_i(t)$ models the small bumps along the slope. We build upon this and consider disturbance of the form of \eqref{equ: disturbance} as ramp disturbances naturally arise in a wide range of applications, see Remark \ref{rmk:disturbance}.%8290989, ramp disturbance
\begin{figure}[thbp]
\centering
\includegraphics[width=\linewidth]{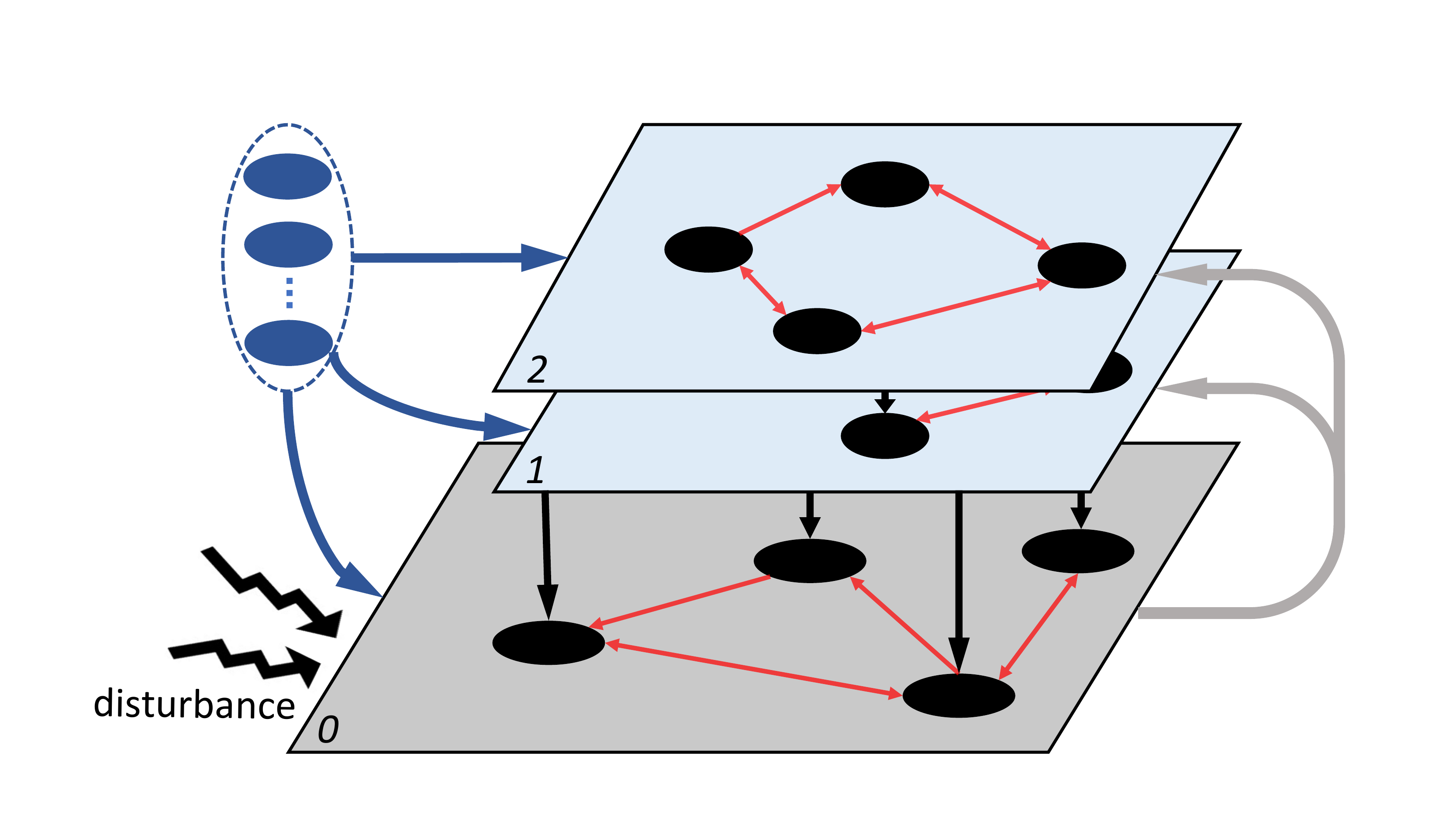}
\caption{Schematic representation of our multiplex architecture.  Black nodes denote the agents while blue nodes represent the (possible) leaders. Colours online.}
\label{fig:multiplex_network}
\end{figure}
Our goal in this paper is to design the control protocol $u_i(t)$ in (\ref{equ: dynamics}) so that the ramp disturbance in (\ref{equ: disturbance}) is rejected, while ensuring a {\em scalability} property of the network system with respect to the residual disturbance $w_i(t)$ (see Section \ref{sec:control_goal} for a rigorous statement of control goal). To do so, we propose the multiplex network architecture schematically shown in Figure \ref{fig:multiplex_network}. In such a figure, the bottom layer (i.e. layer $0$) consists of the network system (\ref{equ: dynamics}) and the multiplex layers (layer $1, 2$) concur to contribute to the control protocol (\ref{equ: control}):
\begin{align}\label{equ: control}
    \begin{split}
        u_i(t)&=h_{i,0}(x(t),x_l(t),t)+h_{i,0}^{(\tau)}(x(t-\tau(t)),x_l(t-\tau(t)),t)\\
        &+r_{i,1}(t),\\
        \dot{r}_{i,1}(t)&=h_{i,1}(x(t),x_l(t),t)+h_{i,1}^{(\tau)}(x(t-\tau(t)),x_l(t-\tau(t)),t)\\
        &+r_{i,2}(t),\\
        \dot{r}_{i,2}(t)&=h_{i,2}(x(t),x_l(t),t)+h_{i,2}^{(\tau)}(x(t-\tau(t)),x_l(t-\tau(t)),t).
    \end{split}
\end{align}
In the above expression, $r_{i,1}(t), r_{i,2}(t)$ are the outputs generated by multiplex layer $1$ and $2$, respectively. As illustrated in Figure \ref{fig:multiplex_network}, the multiplex layers receive information of agents from layer $0$ (grey arrows). Each layer then outputs the signal to the layer immediately below (black arrows), that is layer $2$ outputs $r_{i,2}(t)$ to layer $1$ and layer $1$ outputs $r_{i,1}(t)$ to layer $0$. The functions $h_{i,k}:\R^{nN}\times\R^{nM}\times\R_{\ge 0}\rightarrow\R^n$ and $h_{i,k}^{(\tau)}: \R^{nN}\times \R^{nM}\times\R_{\ge 0} \rightarrow\R^n$, $k=0,1,2$, include both (leader and leaderless) delayed and delay-free couplings (see Remark \ref{rmk:control_examples} for an example). The coupling functions model, on layer $k$ in Figure \ref{fig:multiplex_network}, the connections between the agents (red arrows), either directed or undirected, and the possible links from a group of leaders to the agents (blue arrows). Note that in the case when leaders are present, not all the agents are necessarily connected to them. Also, $x(t)=[x_1\T(t),\ldots,x_N\T(t)]\T$ is the state of the network and $x_l(t)=[x_{l_1}\T(t),\ldots,x_{l_M}\T(t)]\T$ is the state of a group of $M$ leaders. In (\ref{equ: control}) we assume that the delay vector $\tau(t)$ is bounded, i.e.  $\tau(t) \le \tau_{\max}, \forall t$. In what follows, we simply term the smooth coupling functions $h_{i,k}(\cdot,\cdot)$ as {\em delay-free} coupling functions, while the functions $h_{i,k}^{(\tau)}(\cdot,\cdot)$ are termed as {\em delayed} coupling functions. As noted in e.g. \citep{9353260} situations where there is an overlap between delayed and non-delayed communications naturally arise in the context of e.g. platooning, formation control and neural networks. Finally, in (\ref{equ: control}) we set, $\forall s \in [t_0-\tau_{\max}, t_0], \forall i=1,\ldots,N$, $k=1,2$,  $x_i(s)=\varphi_i(s), r_{i,k}(s)=\phi_{i,k}(s)$, with $\varphi_i(s), \phi_{i,k}(s)$ being continuous and bounded functions in $ [t_0-\tau_{\max},t_0]$. 

\begin{remark}\label{rmk:disturbance}
We consider disturbances that consist of a ramp component and a piece-wise continuous component. Ramp disturbances are frequently considered in the literature. See for example \citep{5457987}, where observers for these types of disturbances are considered and \citep{6740883} where the malicious attack is modelled in a close form of \eqref{equ: disturbance}.
\end{remark}
%, 8600548 malicious attack
\begin{remark}\label{rmk:control_examples}
Control protocols of the form of (\ref{equ: control}) arise in a wide range of situations. For example, in the context of formation control typical choices for the coupling functions,  see e.g. \citep{9353260,1261347}, are
\begin{align*}
    \begin{split}
        &h_{i,0}(x(t),x_l(t),t)=\\
        &\sum_{j\in\sN_i}h_{ij}(x_i(t),x_j(t),t)+\sum_{l\in\sL_i}\underline{h}_{il}(x_i(t),x_l(t),t),\\
        &h_{i,0}^{(\tau)}(x(t-\tau(t)),x_l(t-\tau(t)),t)=\\
        &\sum_{j\in\sN_i}h_{ij}^{(\tau)}(x_i(t-\tau(t)),x_j(t-\tau(t)),t)\\
        &+\sum_{l\in\sL_i}\underline{h}_{il}^{(\tau)}(x_i(t-\tau(t)),x_l(t-\tau(t)),t)
    \end{split}
\end{align*}
where $\sN_i$ and $\sL_i$ denote, respectively, the set of neighbours of the $i$-th robot and the set of leaders to which the $i$-th robot is connected. In the above expression, the coupling functions model both delayed and delay-free communications between agents and with the leaders. Typically, these functions are of the diffusive type and no multiplex layers are foreseen in the control architecture.  
\end{remark}
\subsection{Control goal}\label{sec:control_goal}
We let $u(t)=[u_1\T(t),\ldots,u_N\T(t)]\T$ be the stack of the control inputs, $d(t)=[d_1\T(t),\ldots,d_N\T(t)]\T$ be the stack of the disturbances, $w(t)=[w_1\T(t),\ldots,w_N\T(t)]\T$ be the stack of the residual disturbances and $\bar{d}(t)=[\bar d_1\T(t),\ldots,\bar d_N\T(t)]\T$ be the stack of the ramp disturbances. We also let $\bar{d}_0=[\bar{d}_{1,0},\ldots,\bar{d}_{N,0}]\T$ and $\bar{d}_1=[\bar{d}_{1,1},\ldots,\bar{d}_{N,1}]\T$. Our control goal is expressed in terms of the so-called desired solution of the disturbance-free (or unperturbed in what follows) network system following \citep{MONTEIL2019198}. Intuitively, the desired solution is the solution of the network system characterized by having: (i) the state of the agents attaining some desired configuration; (ii) the multiplex layers giving no contribution to the $u_i$'s. Formally, the desired solution is the solution of network system (\ref{equ: dynamics}) controlled by (\ref{equ: control}) such that: (i) $x^\ast(t):=[x_1^{\ast\mathsf{T}}(t), \dots,x_N^{\ast\mathsf{T}}(t)]\T$, with $\dot{x}^\ast_i(t)=f_i(x_i^\ast(t),t), \forall i$; (ii) $r_{i,k}^{\ast}(t) = 0$, $\forall i,\forall k$ and $\forall t$. It is intrinsic in this definition that when the desired solution is achieved it must hold that $u_i(t) = 0$ (note that this property is satisfied by e.g. any diffusive-type control protocol). In what follows, for the sake of brevity, we make a slight abuse of terminology and say $x^\ast(t)$ is desired solution.\\
We aim at designing the control protocol (\ref{equ: control}) so that the closed loop system rejects the ramp disturbances while guaranteeing that the residual disturbances $w(t)$ are not amplified within the network system. This is captured by the definition of scalability with respect to $w(t)$ formalized next:
\begin{definition}\label{def: L_inf}
Consider the closed loop system (\ref{equ: dynamics}) - (\ref{equ: control}) with disturbance $d(t)=w(t) + \bar{d}(t)$. The system is $\sL_\infty^p$-Input-to-State Scalable with respect to $w(t)$ if there exists class $\sKL$ functions $\alpha(\cdot,\cdot)$, $\beta(\cdot,\cdot)$, a class $\sK$ function $\gamma(\cdot)$, such that for any initial condition and $\forall t \ge t_0$, 
    \begin{align*}
        \begin{split}
            &\max_i\abs{x_i(t)-x_i^\ast(t)}_p \le\\
            &\alpha\left(\max_i\sup_{t_0-\tau_{\max}\le s \le t_0}\abs{x_i(s)-x_i^\ast(s)}_p,t-t_0\right)\\
            &+\beta\Biggl(\max_i\sup_{t_0-\tau_{\max}\le s \le t_0}\Bigl(\abs{r_{i,1}(s)+\bar{d}_{i,0}+\bar{d}_{i,1} \cdot s}_p\\
            &+\abs{r_{i,2}(s)+\bar{d}_{i,1}}_p\Bigr),t-t_0\Biggr)+\gamma\left(\max_i\norm{w_i(\cdot)}_{\sL_\infty^p}\right), \forall N.
        \end{split}
    \end{align*}
    %\item $\sL_\infty^p$-Input-Output Scalable with respect to $w(t)$: if there exists class $\sKL$ functions $\alpha(\cdot,\cdot)$, $\beta(\cdot,\cdot)$, a class $\sK$ function $\gamma(\cdot)$, such that for any initial condition and $\forall t \ge t_0$, 
    %\begin{align*}
    %    \begin{split}
            %&\max_i\abs{y_i(t)-y_i^*(t)}_p \le\\
            %&\alpha\left(\max_i\sup_{t_0-\tau_{\max}\le s \le t_0}\abs{x_i(s)-x_i^*(s)}_p,t-t_0\right)\\
            %&+\beta\Biggl(\max_i\sup_{t_0-\tau_{\max}\le s \le t_0}\Bigl(\abs{r_{i,0}(s)+\bar{d}_{i,0}+\bar{d}_{i,1} \cdot s}_p\\
            %&+\abs{r_{i,1}(s)+\bar{d}_{i,1}}_p\Bigr),t-t_0\Biggr)+\gamma\left(\max_i\norm{w_i(\cdot)}_{\sL_\infty^p}\right), \forall N.
        %\end{split}
%    \end{align*}
%\end{itemize}
\end{definition}
In the special case when $\bar{d}(t) =0$ and there are no multiplex layers, i.e. $r_{i,k}(t)=0, \forall k$, Definition \ref{def: L_inf}  becomes the definition for scalability given in \citep{9353260}. In this context we note that the bounds in Definition \ref{def: L_inf} are uniform in $N$ and this in turn guarantees that the residual disturbances are not amplified within the network system. In what follows, whenever it is clear from the context, we simply say that the network system is $\sL_\infty^p$-Input-to-State Scalable if Definition \ref{def: L_inf} is fulfilled. In a special case when $p=2$, we use $\sL_\infty$-Input-to-State Scalable for simplicity.
\begin{remark}\label{rem:norm}
With our technical results, we give conditions on the control protocol that ensure scalability of the closed loop system. Essentially, these conditions guarantee a contractivity property of the network system using G-norm $\abs{x}_G=\left\vert\abs{x_1}_p,\ldots,\abs{x_N}_p \right\vert_{\infty}$. 
\end{remark}

\begin{figure*}[b]
\par\noindent\rule{\textwidth}{0.4pt}
\begin{equation}\label{equ: matrix}
\begin{gathered}
    \bar{A}_{ii}(t)=\left[\begin{matrix} \frac{\partial f_i(x_i,t)}{\partial x_i}+\frac{\partial 				h_{i,0}(x,x_l,t)}{\partial x_i} & I_n &  0_n\\
     			\frac{\partial h_{i,1}(x,x_l,t)}{\partial x_i} & 0_n & I_n\\ 
     			\frac{\partial h_{i,2}(x,x_l,t)}{\partial x_i} & 0_n & 0_n 						\end{matrix}\right],  
    \bar{A}_{ij}(t)=\left[\begin{matrix} \frac{\partial h_{i,0}(x,x_l,t)}{\partial x_j} & 0_n & 0_n\\
    \frac{\partial h_{i,1}(x,x_l,t)}{\partial x_j} & 0_n & 0_n \\
      			\frac{\partial h_{i,2}(x,x_l,t)}{\partial x_j} & 0_n & 0_n 						\end{matrix}\right],
    \bar{B}_{ij}(t)=\left[\begin{matrix} \frac{\partial h_{i,0}^{(\tau)}(x,x_l,t)}{\partial x_j} & 					0_n & 0_n\\
      			\frac{\partial h_{i,1}^{(\tau)}(x,x_l,t)}{\partial x_j} & 0_n & 0_n \\
      			\frac{\partial h_{i,2}^{(\tau)}(x,x_l,t)}{\partial x_j} & 0_n  & 0_n \end{matrix}\right]
\end{gathered}
\end{equation}
\end{figure*}
\section{Technical result}\label{sec: technical}
We now introduce our main technical result. For the network system (\ref{equ: dynamics}) we give sufficient conditions on the control protocol (\ref{equ: control}) guaranteeing that the closed-loop system affected by disturbances of the form (\ref{equ: disturbance}) is $\sL_\infty^p$-Input-to-State Scalable (see Definition \ref{def: L_inf}). The results are stated in terms of the block diagonal state transformation matrix $T:=\textbf{diag}\{T_1,\ldots,T_N\}\in \R^{3nN \times 3nN}$ with
$$T_i:=\left[\begin{matrix}I_{n} & \alpha_{i,1}\cdot I_{n} &  0_n\\
0_n & I_{n} & \alpha_{i,2}\cdot I_n\\
0_n & 0_n & I_{n} \end{matrix}\right] \in \R^{3n \times 3n},$$
where $\alpha_{i,1}, \alpha_{i,2} \in \R$.
%\GR{TBD: as discussed, check everything now is fine with C1, the desired solution etc...I also made the edits to the control goal section. Also, move the equations in (4) on the same page as Proposition 1 (perhaps we can fix this when we have the final formatting of the paper - just take a note for now)} with $y_i(t) = x_i(t)$
\begin{proposition}\label{prop: scalability}
Consider the closed-loop network system (\ref{equ: dynamics}) with control protocol (\ref{equ: control}) affected by disturbances (\ref{equ: disturbance}).  Assume that, $\forall t\ge t_0$, the following set of conditions are satisfied for some $0\le\underline{\sigma}<\bar{\sigma}<+\infty$:
\begin{itemize}
    \item[C1] $h_{i,k}(x^\ast,x_l,t)=h_{i,k}^{(\tau)}(x^\ast,x_l,t)=0$, $\forall i, k$;
    \item[C2] $\mu_p(T_i\bar{A}_{ii}(t)T_i^{-1})+\sum_{j\ne i}\norm{T_i\bar{A}_{ij}(t)T_j^{-1}}_p\le -\bar{\sigma}$,  $\forall i$ and {$\forall x\in\R^{nN}, \forall x_l \in \R^{nM}$} (the state dependent matrices $\bar{A}_{ij}(t)$'s are defined in (\ref{equ: matrix}));
    \item[C3] $\sum_j\norm{T_i\bar{B}_{ij}(t)T_j^{-1}}_p \le \underline{\sigma}$, $ \forall i$ and {$\forall x\in\R^{nN}, \forall x_l \in \R^{nM}$} (the state dependent matrices $\bar{B}_{ij}(t)$'s are also defined in (\ref{equ: matrix})).
\end{itemize}
Then, the system is $\sL_\infty^p$-Input-to-State Scalable with
\begin{align}\label{eqn:upper_bound_desired}
\begin{split}
        &\max_i\abs{x_i(t)-x_i^\ast(t)}_p \le \\
        &\kappa_G(T)e^{-\hat\lambda (t-t_0)}\max_i\sup_{t_0-\tau_{\max}\le s \le t_0}\abs{x_i(s)-x_i^\ast(s)}_p\\
        &+\kappa_G(T)e^{-\hat\lambda (t-t_0)}\max_i\sup_{t_0-\tau_{\max}\le s \le t_0}\Bigl(\abs{r_{i,2}(s)+\bar{d}_{i,1}}_p\\
        &+\abs{r_{i,1}(s)+\bar{d}_{i,0}+\bar{d}_{i,1}\cdot s}_p\Bigr)+\frac{\kappa_G(T)}{\bar{\sigma}-\underline{\sigma}}\max_i \norm{w_i(\cdot)}_{\sL_\infty^p}, \forall N,
\end{split}
\end{align}
where $\kappa_G(T):=\norm{T}_G\norm{T^{-1}}_G$, $\hat\lambda=\inf_{t\ge t_0}\{\lambda|\lambda(t)-\bar{\sigma}+\underline{\sigma}e^{\lambda(t)\tau(t)}=0\}$, $x_i(t)$ is a solution of agent $i$ with $x_i(s)=\varphi_i(s), r_{i,k}(s)=\phi_{i,k}(s)$ and $x_i^\ast(s)=x_i^\ast(t_0)$, $s\in [t_0-\tau_{\max},t_0], i=1,\ldots, N, k=1,2$.
\end{proposition}
\proof
We start with augmenting the state of the original dynamics by defining $z_i(t):=[x_i\T(t), \zeta_{i,1}\T(t), \zeta_{i,2}\T(t)]\T$, and where
\begin{align*}
    \begin{split}
        \zeta_{i,1}(t)&=r_{i,1}(t) + \bar{d}_{i,0} + \bar{d}_{i,1}\cdot t,\\
        \zeta_{i,2}(t)&=r_{i,2}(t) + \bar{d}_{i,1}.
    \end{split}
\end{align*}
In these new coordinates the dynamics of the network system becomes
\begin{align*}
    \dot{z}_i(t)=\phi_i(z_i,t)+v_i(z,t)+ \tilde w_i(t),
\end{align*}
where $\phi_i(z_i,t) =[f_i\T(x_i,t),0_{1\times n},0_{1\times n}]\T$, $\tilde w_i(t)=[w_i\T(t),0_{1\times n}, \allowbreak 0_{1\times n}]\T$, and where 
\begin{align*}
    \begin{split}
        &v_i(z,t)  = \\
&\left[\begin{array}{*{20}c}
h_{i,0}(x(t),x_l(t),t) + h_{i,0}^{(\tau)}(x(t-\tau(t)),x_l(t-\tau(t)),t) + \zeta_{i,1}(t)\\
h_{i,1}(x(t),x_l(t),t) + h_{i,1}^{(\tau)}(x(t-\tau(t)),x_l(t-\tau(t)),t) + \zeta_{i,2}(t)\\
h_{i,2}(x(t),x_l(t),t) + h_{i,2}^{(\tau)}(x(t-\tau(t)), x_l(t-\tau(t)),t)\end{array}\right].
    \end{split}
\end{align*}
Note that $C1$ implies that the desired configuration $x_i^\ast(t)$ is a solution of the unperturbed network dynamics, i.e. $x_i^\ast(t)$ satisfies $\dot x_i^\ast(t)=f_i(x_i^\ast,t)$. Moreover, {when there are no disturbances,} in the new set of coordinates, the solution $z_i^\ast(t):=[x_i^{\ast\mathsf{T}}(t), 0_{1\times n}, 0_{1\times n}]\T$ satisfies
\begin{align*}
    \dot{z}^\ast_i(t)=\phi_i(z_i^\ast,t),
\end{align*}
with $\phi_i(z_i^\ast,t)=[f_i\T(x_i^\ast,t),0_{1\times n}, 0_{1\times n}]\T$. Hence, the dynamics of state deviation $e_i(t)=z_i(t)-z_i^\ast(t)$ is given by
\begin{align*}
    \begin{split}
        \dot{e}_i(t)&=\phi_i(z_i,t)-\phi_i(z_i^\ast,t)+v_i(z,t)+\tilde w_i(t).
    \end{split}
\end{align*}
Following e.g. \citep{1083507}, we let $\eta_i(\rho)=\rho z_i+(1-\rho)z_i^\ast$ and $\eta(\rho)=[\eta_1\T(\rho),\ldots,\eta_N\T(\rho)]\T$ and then rewrite the error dynamics as
\begin{align*}
    \dot{e}(t)=A(t)e(t)+B(t)e(t-\tau(t))+\tilde w(t),
\end{align*}
where $\tilde w=[\tilde w_1\T(t),\ldots,\tilde w_N\T(t)]\T$ and $A(t)$ has entries: (i) $A_{ii}(t)=\int_0^1 (J_{\phi_i}(\eta_i(\rho),t)+J_{v_i}^{(0)}(\eta_i(\rho),t))d\rho$; (ii) $A_{ij}(t)=\int_0^1 J_{v_i}^{(0)}(\eta_j(\rho),t)d\rho$. Similarly, $B(t)$ has entries: $B_{ij}(t)=\int_0^1 J_{v_i}^{(\tau)}(\eta_j(\rho),t)d\rho$. In the above expressions, the Jacobian matrices are defined as $J_{\phi_i}(\eta_i,t) := \frac{d \phi_i(\eta_i,t)}{d \eta_i}$, $J_{v_i}^{(0)}(\eta_i,t) := \frac{d v_i^{(0)}(\eta,t)}{d \eta_i}$, $J_{v_i}^{(\tau)}(\eta_i,t) :=\frac{d v_i^{(\tau)}(\eta,t)}{d \eta_i}$ where the superscripts $(0)$ and $(\tau)$ denote the delay-free and the delayed components of $v_i$, respectively. Now, let $\tilde{z}(t) :=Tz(t)$ and  $\tilde{e}(t):=Te(t)$. Then, we have
\begin{align*}
    \dot{\tilde{e}}(t)=TA(t)T^{-1}\tilde{e}(t)+TB(t)T^{-1}\tilde{e}(t-\tau(t))+T\tilde w(t).
\end{align*}
Also, by taking the Dini derivative of $\abs{\tilde{e}(t)}_{G}$ we obtain
\begin{align*}
    \begin{split}
        &D^+\abs{\tilde{e}(t)}_{G} \le \mu_{G}(TA(t)T^{-1})\abs{\tilde{e}(t)}_{G} +\norm{T}_G\max_i \norm{\tilde w_i(\cdot)}_{\sL_\infty^p}\\
        &+ \norm{TB(t)T^{-1}}_{G} \sup_{t-\tau_{\max}\le s \le t}\abs{\tilde e(s)}_{G}.
    \end{split}
\end{align*}
Next, we find upper bounds for $\mu_{G}(TA(t)T^{-1})$ and $\norm{TB(t)T^{-1}}_{G}$ which allow us to apply Lemma \ref{prop:halanay}. First, we give the expression of the matrix $\bar{A}(t)$ which have entries defined in \eqref{equ: matrix}: (i) $\bar{A}_{ii}(t)=J_{\phi_i}(z_i,t)+J_{v_i}^{(0)}(z_i,t)$; (ii) $\bar{A}_{ij}(t)=J_{v_i}^{(0)}(z_j,t)$ and $\bar{B}(t)$ has entries: $\bar{B}_{ij}(t)=J_{v_i}^{(\tau)}(z_j,t)$. Then, by sub-additivity of matrix measures and matrix norms, we get
$\mu_{G}(TA(t)T^{-1}) \le \int_0^1 \mu_{G}(T \bar A(t)T^{-1})d\rho$ and $\norm{TB(t)T^{-1}}_{G} \le \int_{0}^1\norm{T\bar B(t)T^{-1}}_{G}d\rho$ (see also Lemma $3.4$ in \citep{1531-3492_2021188}). Moreover, from Lemma \ref{lem:matrix norm} it then follows that
\begin{align*}
    &\mu_{G}(T\bar A(t)T^{-1}) \le\\
    &\max_i\left\{\mu_p(T_i \bar A_{ii}(t)T_i^{-1}) +\sum_{j\ne i}\norm{T_i \bar A_{ij}(t)T_j^{-1}}_p\right\}
\end{align*}
and 
\begin{align*}
    \norm{T\bar B(t)T^{-1}}_{G} \le \max_i\left\{\sum_j\norm{T_i \bar B_{ij}(t)T_j^{-1}}_p\right\}.
\end{align*}
Condition $C2$ and $C3$ yields that 
\begin{align*}
    \max_i\left\{\mu_p(T_i \bar A_{ii}(t)T_i^{-1})+\sum_{j\ne i}\norm{T_i \bar A_{ij}(t)T_j^{-1}}_p\right\} \le -\bar{\sigma}
\end{align*}
and
\begin{align*}
    \max_i\left\{\sum_j\norm{T_i \bar B_{ij}(t)T_j^{-1}}_p\right\} \le \underline{\sigma}   
\end{align*}
for some $0\le\underline{\sigma}<\bar{\sigma}<+\infty$. This implies that $\mu_G(TA(t)T^{-1})+\norm{TB(t)T^{-1}}_G\le \bar{\sigma} + \underline{\sigma}:= -\sigma$ and Lemma \ref{prop:halanay} then yields 
\begin{align*}
    \abs{\tilde{e}(t)}_{G}\le \sup_{t_0-\tau_{\max}\le s \le t_0}\abs{\tilde{e}(s)}_{G} e^{-\hat\lambda (t-t_0)} + \frac{\norm{T}_G}{\bar{\sigma}-\underline{\sigma}}\max_i \norm{\tilde w_i(\cdot)}_{\sL_\infty^p},
\end{align*}
with $\hat\lambda$ defined as in the statement of the proposition. Since $\tilde{e}=Te$ we get $\abs{e(t)}_G \le \norm{T^{-1}}_G\abs{\tilde{e}(t)}_G$ and $\abs{\tilde{e}(t)}_G \le \norm{T^{-1}}_G\abs{e(t)}_G$. We also notice that the definition of $\tilde w_i(\cdot)$ implies that $\norm{\tilde w_i(\cdot)}_{\sL_\infty^p}=\norm{w_i(\cdot)}_{\sL_\infty^p}$. Hence
\begin{align*}
    \abs{e(t)}_G  \le & \norm{T^{-1}}_G\norm{T}_G\sup_{t_0-\tau_{\max}\le s \le t_0}\abs{e(s)}_{G} e^{-\hat\lambda (t-t_0)} \\
    &+ \frac{\norm{T^{-1}}_G\norm{T}_G}{\bar{\sigma}-\underline{\sigma}}\max_i \norm{w_i(\cdot)}_{\sL_\infty^p}.
\end{align*}
We note that 
\begin{align*}
    \begin{split}
        &\abs{e_i(t)}_p=\left\vert\left[\begin{matrix}
    x_i(t)-x_i^\ast(t) \\ \zeta_{i,1}(t)\\ \zeta_{i,2}(t) 
    \end{matrix}\right]\right\vert_p \ge \left\vert\left[\begin{matrix}
    x_i(t)-x_i^\ast(t) \\ 0_{n\times 1} \\ 0_{n\times 1}
    \end{matrix}\right]\right\vert_p,\\
        &\abs{e_i(t)}_p \le \left\vert\left[\begin{matrix}
    x_i(t)-x_i^\ast(t) \\ 0_{n \times 1} \\ 0_{n\times 1}
    \end{matrix}\right]\right\vert_p + \left\vert\left[\begin{matrix}
    0_{n\times 1} \\ \zeta_{i,1}(t) \\ 0_{n\times 1}
    \end{matrix}\right]\right\vert_p  + \left\vert\left[\begin{matrix}
    0_{n\times 1} \\ 0_{n\times 1} \\ \zeta_{i,2}(t)
    \end{matrix}\right]\right\vert_p.    
    \end{split}
\end{align*}
Hence, $\abs{x_i(t)-x_i^\ast(t)}_p \le \abs{e_i(t)}_p$ and $\abs{e_i(t)}_p \le \abs{x_i(t)-x_i^\ast(t)}_p+ \abs{\zeta_{i,1}(t)}_p+\abs{\zeta_{i,2}(t)}_p$. We then finally obtain the upper bound of the state deviation
\begin{align*}
\begin{split}
        &\max_i\abs{x_i(t)-x_i^\ast(t)}_p \le \\
        &\kappa_G(T)e^{-\hat\lambda (t-t_0)}\max_i\sup_{t_0-\tau_{\max}\le s \le t_0}\abs{x_i(s)-x_i^\ast(s)}_p\\
        &+\kappa_G(T)e^{-\hat\lambda (t-t_0)}\max_i\sup_{t_0-\tau_{\max}\le s \le t_0}\Bigl(\abs{r_{i,2}(s)+\bar{d}_{i,1}}_p\\
        &+\abs{r_{i,1}(s)+\bar{d}_{i,0}+\bar{d}_{i,1}\cdot s}_p\Bigr)+\frac{\kappa_G(T)}{\bar{\sigma}-\underline{\sigma}}\max_i \norm{w_i(\cdot)}_{\sL_\infty^p}, \forall N.
\end{split}
\end{align*}
\endproof

\begin{remark}
$C1$ implies that $u_i(t)=0$ at the desired solution. This rather common condition (see e.g. \citep{6891349,9353260}) guarantees that $x^\ast(t)$ is a solution of the unperturbed dynamics. $C2$ gives an upper bound on matrix measure of the Jacobian of the intrinsic dynamics and of the delay-free part of the protocol. $C2$ says that such a matrix measure should be negative enough to compensate for the norm of the Jacobian of the delayed part of the protocol, whose upper bound is given in  $C3$. 
\end{remark}
%%%% SUGGEST REMOVING THIS AS IT DOES NOT SEEM TO ADD CLARITY (IF YOU WANT, HOWEVER, YOU CAN KEEP THE REMARK %%%
%\begin{remark}
%$C1$ is easy to satisfy in the design stage of the control protocol. To fulfill $C2$, $C3$, in a practical sense, a solution is to find the upper bounds for the involved matrix measure and matrix norms separately, which depend on the choice of the coupling functions. The transformation matrices can also be selected to be uniform for simplicity. More details on implementing the conditions in practice will not be presented here for the sake of briefness.
%\end{remark}
\begin{remark}
The {\em convergence rate} $\hat\lambda$ depends on $\tau_{\max}$. Indeed, for a given set of fixed $\bar{\sigma}$, $\underline{\sigma}$, increasing $\tau_{\max}$ decreases $\hat{\lambda}$. Hence, a trade-off exists between the length of delay and the convergence rate. In particular, $\hat \lambda=\bar{\sigma}$ when $\tau_\max=0$ which yields the results in \citep{MONTEIL2019198} for a delay-free  platoon system with no multiplex layers.
\end{remark}

%The next result immediately follows from Proposition \ref{prop: scalability}.
%\begin{corollary}
%consider the closed loop network system (\ref{equ: dynamics}) - (\ref{equ: control}) affected by disturbances (\ref{equ: disturbance}). Assume that all the conditions in Proposition \ref{prop: scalability} are satisfied and that, in addition, the output functions $g_i(\cdot)$ are Lipschitz. Then, (\ref{equ: dynamics}) - (\ref{equ: control}) is $\sL_\infty^p$-Input-Output Scalable.
%\end{corollary}
%\proof The proof is omitted here for brevity. \qed

\section{Application}\label{sec: application}
We show the effectiveness of the result by designing a control protocol satisfying the conditions in Proposition \ref{prop: scalability} so that a network of $N$ unicycle robots is $\mathcal{L}_\infty$-Input-to-State Scalable. In particular, we aim at designing a 
formation where local residual disturbances on one robot are not amplified and the robots in the formation are required to (i) track the reference provided by a virtual leader; (ii) reject polynomial disturbances up to ramps. 

\subsubsection*{Unicycle dynamics.}
We consider the following dynamics
\begin{align}\label{equ: unicycle_dynamics}
    \begin{split}
        \dot{p}_i^x&=v_i\cos\theta_i+d_i^x(t),\\
        \dot{p}_i^y&=v_i\sin\theta_i+d_i^y(t),\\
		\dot{\theta}_i&=\Omega_i,
    \end{split}
\end{align}
$\forall i$, where the state variables $p_i(t)=[p_i^x(t),p_i^y(t)]\T$ is the inertial position and $\theta_i(t)$ is the robot heading angle. The control input is denoted as $u_i(t)=[v_i(t),\Omega_i(t)]\T$ with $v_i(t)$ being the linear velocity and $\Omega_i(t)$ being the angular velocity. The disturbances affecting the robots are $d_i(t)=[d_i^x(t),d_i^y(t)]\T$ where  $d_i^x(t):=\bar{d}_{i,0}^{x}+\bar{d}_{i,1}^{x}\cdot t+w_i^x(t)$ and $d_i^y(t):=\bar{d}_{i,0}^{y}+\bar{d}_{i,1}^{y}\cdot t+w_i^y(t)$. We introduce the following compact notation: $w_i(t)=[w_i^x(t),w_i^y(t)]\T$,  $\bar{d}_i(t)=[\bar{d}_{i,0}^{x}, \bar{d}_{i,0}^{y}]\T+[\bar{d}_{i,1}^x\cdot t, \bar{d}_{i,1}^y\cdot t]\T$. The constant term $[\bar{d}_{i,0}^x, \bar{d}_{i,0}^y]$ can model, for example in the case of unicycle-like marine robots, the constant speed disturbance caused by the ocean current \citep{7463532} and the residual term $w_i(t)$ models the transient variation of the current. The ramp term $[\bar{d}_{i,1}^x\cdot t, \bar{d}_{i,1}^y\cdot t]$ can model e.g. ramp attack signals \citep{6740883}. Following \citep{1261347}, the dynamics for the robot hand position is given by
\begin{align}\label{equ: fbl_dynamics}
	\dot{\eta}_i(t)=\left[\begin{matrix} \cos\theta_i & -l_i\sin\theta_i \\ \sin\theta_i & l_i\cos\theta_i\end{matrix}\right]u_i(t)+d_i(t),
\end{align}
where $l_i\in \R_{>0}$ is the distance of the hand position to the wheel axis. The dynamics can be feedback linearised by
\begin{align*}
    u_i(t)=\left[\begin{matrix} \cos\theta_i & -l_i\sin\theta_i \\ \sin\theta_i & l_i\cos\theta_i\end{matrix}\right]^{-1}\nu_i(t),
\end{align*}
which yields
\begin{align}\label{equ: dynamics_hand_position}
	\dot{\eta}_i(t)=\nu_i(t)+d_i(t), \ \ \ \forall i
\end{align}
Next we leverage Proposition \ref{prop: scalability} to design $\nu_i(t)$ so that network \eqref{equ: dynamics_hand_position} is $\sL_\infty$-Input-to-State Scalable.

\subsubsection*{Protocol design.}
\begin{figure}[thbp]
\centering
\includegraphics[width=\columnwidth]{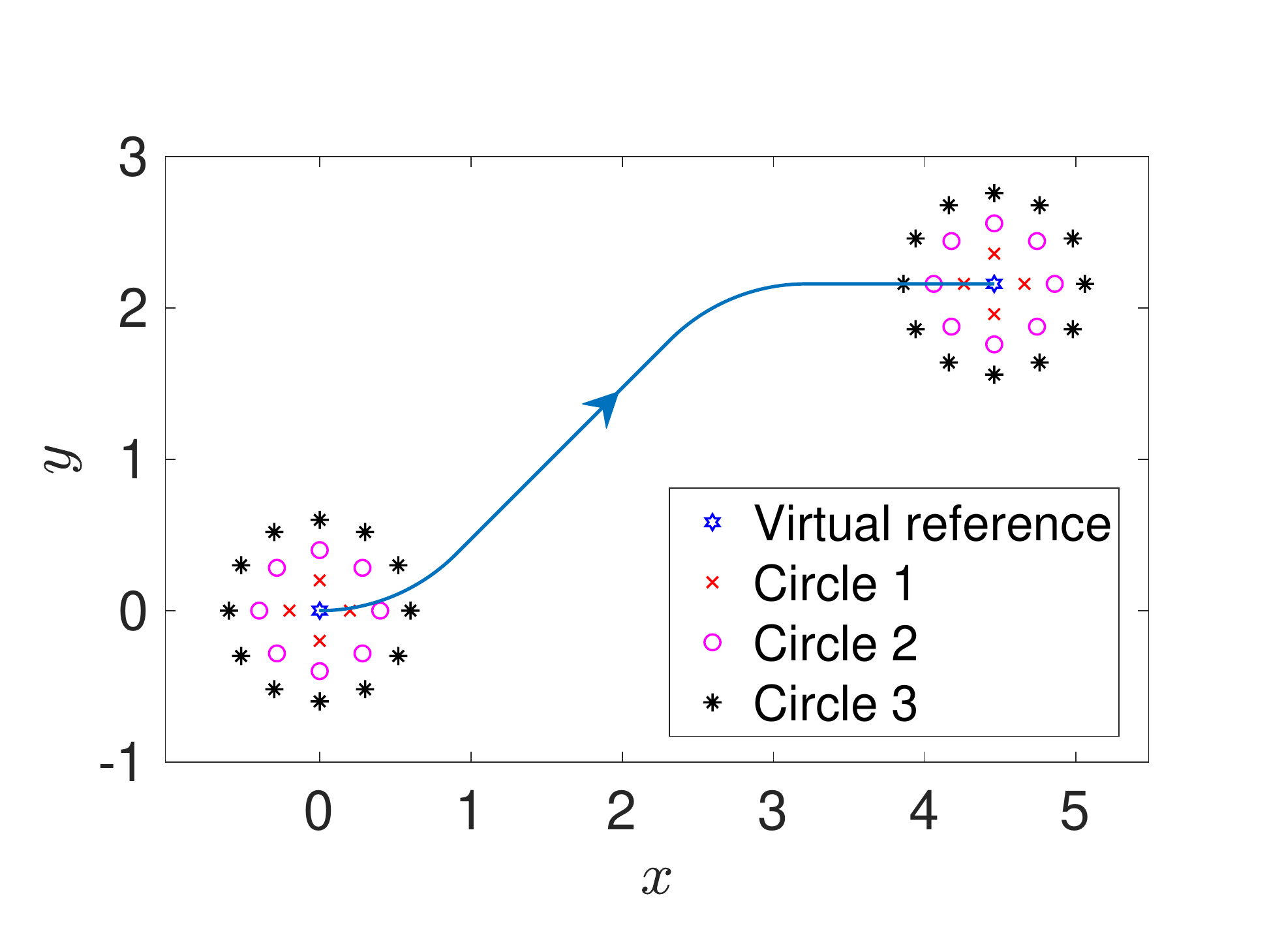}
\caption{Reference trajectory of the hand position from the virtual leader and an example of  desired formation.}
\label{fig:formation_pattern}
\end{figure}

We denote by $\eta_l$ the hand position provided by a virtual leader. Robots are required to keep a desired offset from the leader ($\delta_{li}^\ast$) and from neighbours ($\delta_{ji}^\ast$) while tracking a reference speed from the leader, say $v_l$. Hence, the desired position of the $i$-th robot, $\eta_i^\ast$, is picked so that: (i) the robot keeps the desired offsets from the leader and from the neighbours, i.e. $\eta_l^\ast-\eta_i^\ast=\delta_{li}^\ast$ and $\eta_j^\ast-\eta_i^\ast=\delta_{ji}^\ast$; (ii) the reference speed is tracked, i.e. $\dot{\eta}_i^\ast=v_l$. To this aim, we propose the following control protocol 
\begin{align}\label{equ: fbl_control}
	\begin{split}
		\nu_i(t)&=h_{i,0}(\eta(t),\eta_l(t),t)+h_{i,0}^{(\tau)}(\eta(t-\tau(t)),\eta_l(t-\tau(t)),t)\\
		&+v_l(t)+r_{i,1}(t),\\
        \dot{r}_{i,1}(t)&=h_{i,1}(\eta(t),\eta_l(t),t)+h_{i,1}^{(\tau)}(\eta(t-\tau(t)),\eta_l(t-\tau(t)),t)\\
        &+r_{i,2}(t),\\
        \dot{r}_{i,2}(t)&=h_{i,2}(\eta(t),\eta_l(t),t)+h_{i,2}^{(\tau)}(\eta(t-\tau(t)),\eta_l(t-\tau(t)),t),
	\end{split}
\end{align}
where the coupling functions $h_{i,k}, h_{i,k}^{(\tau)}: \R^{2N}\times \R^2 \times \R_{\ge 0} \rightarrow \R^2$ are smooth functions for delay-free and delayed couplings between the robots and the leader of the form:
\begin{align}\label{equ: fbl_coupling}
	\begin{split}
		h_{i,0}(\eta(t),\eta_l(t),t)&=k_{0}(\eta_l-\eta_i-\delta_{li}^\ast),\\		h_{i,0}^{(\tau)}(\eta(t),\eta_l(t),t)&=k_{0}^{(\tau)}\sum_{j\in \sN_i}\psi(\eta_j-\eta_i-\delta_{ji}^\ast),\\
        h_{i,1}(\eta(t),\eta_l(t),t)&=k_{1}(\eta_l-\eta_i-\delta_{li}^\ast),\\ 
		h_{i,1}^{(\tau)}(\eta(t),\eta_l(t),t)&=k_{1}^{(\tau)}\sum_{j\in \sN_i}\psi(\eta_j-\eta_i-\delta_{ji}^\ast),\\
        h_{i,2}(\eta(t),\eta_l(t),t)&=k_{2}(\eta_l-\eta_i-\delta_{li}^\ast),\\ 
		h_{i,2}^{(\tau)}(\eta(t),\eta_l(t),t)&=k_{2}^{(\tau)}\sum_{j\in \sN_i}\psi(\eta_j-\eta_i-\delta_{ji}^\ast),
	\end{split}
\end{align}
with $\psi(x):=\tanh(k^{\psi}x)$ inspired from \citep{MONTEIL2019198}. In the above expression, $\sN_i$ is the set of neighbours that robot $i$ is connected to and its cardinality is bounded, i.e. $\textbf{card}(\sN_i)\le\bar{N}, \forall i$. The control gains $k_{0},$ $k_{1},$ $k_{2},$ $k_{0}^{(\tau)}, k_{1}^{(\tau)}, k_{2}^{(\tau)}, k^{\psi}$ are positive scalars designed next. The desired formation consists of concentric circles with the $k$-th circle having $4k$ robots. A robot on the $k$-th circle is connected to at most $\bar{N}=3$ other robots, i.e. the ones immediately ahead and behind on the same circle and the closest robot on circle $k-1$ (if any). An example of the desired formation with $3$ concentric circles is shown in Figure \ref{fig:formation_pattern} where the reference trajectory is also plotted.
Next we make use of Proposition \ref{prop: scalability} to select the control gains so that the robotic network is $\sL_\infty$-Input-to-State Scalable. In particular, we note that the choice of the control protocol (\ref{equ: fbl_control}) with coupling functions (\ref{equ: fbl_coupling}) guarantees the fulfillment of $C1$. We then select the set of control gains satisfying condition $C2$ and $C3$ following steps (details omitted for brevity) similar in spirit to \citep{MONTEIL2019198}. This resulted in: $k_{0}=1.4342, k_{1}=1.536, k_{2}=0.4937, k_{0}^{(\tau)}=0.321, k_{1}^{(\tau)}=0.436, k_{2}^{(\tau)}=0.213, k^\psi=0.1$. 

\subsubsection*{Numerical validation.}
\begin{figure}[thbp]
\centering
\includegraphics[width=\columnwidth]{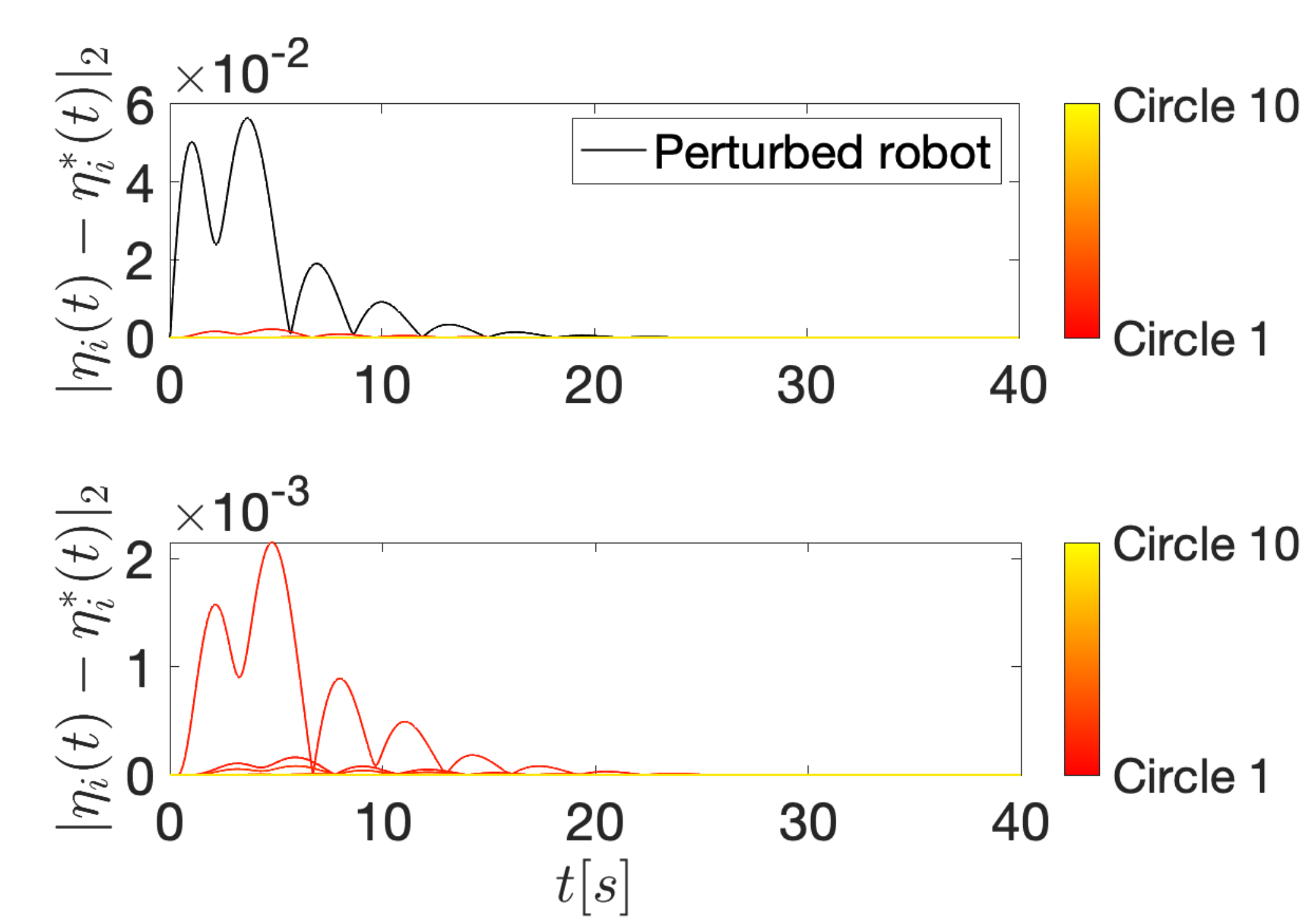}
\caption{Top panel -- Robots hand position deviations (in meters) when one robot in circle $1$ is affected by disturbance $d_{i}(t)$; Bottom panel -- Hand position deviations for robots not directly affected by the disturbance only. Robots on the same circle have same color except the perturbed one. Colours online.}
\label{fig:hand_position_deviation}
\end{figure}

We validate the effectiveness of the control protocol \eqref{equ: fbl_control} designed above by illustrating that: (i) the robots achieve the desired formation, while following the reference trajectory; (ii) polynomial disturbances up to ramps are rejected; (iii) the local residual disturbances on one robot are not amplified. In the simulation, we consider a formation of $10$ circles where the (hand position of the) robots move at a constant linear speed and one robot in circle $1$ is affected by the disturbance 
\begin{align}\label{equ: numerical_disturbance}
    d_i(t)=\left[\begin{matrix}0.07+0.02t+0.05\sin(t)e^{-0.3t}\\0.06-0.04t+0.06\sin(t)e^{-0.3t}\end{matrix}\right], 
\end{align}
where $w_i(t)=[0.05\sin(t)e^{-0.3t},0.06\sin(t)e^{-0.3t}]\T$ is the residual disturbance signal. The delay is set to $\tau(t)=0.33s$. 
\begin{figure}
         \centering
         \includegraphics[width=\columnwidth]{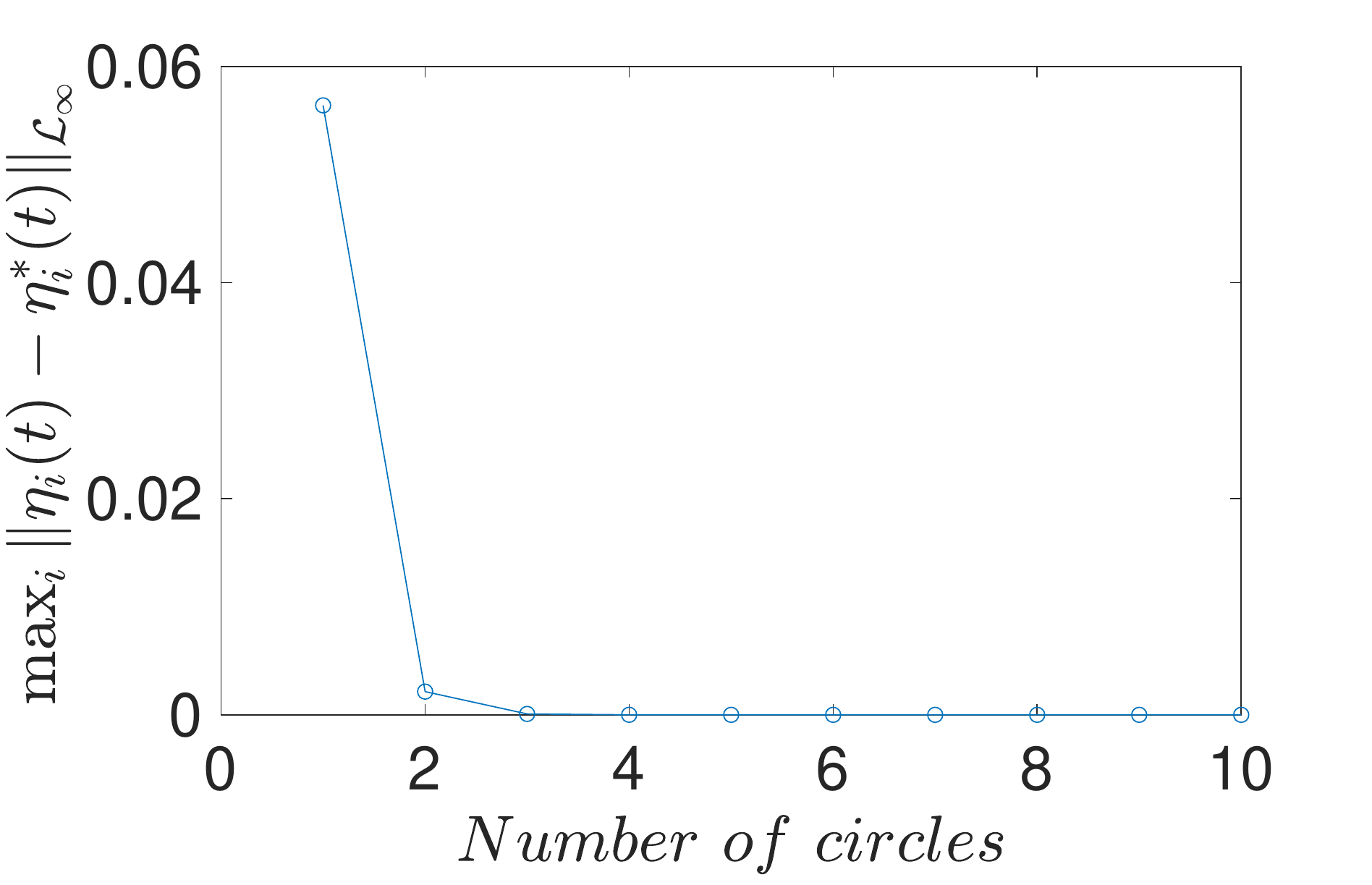}
        \caption{Maximum hand position deviation from the desired position as the number of circles increases.}
        \label{fig:2layer_network}
\end{figure}
Figure \ref{fig:2layer_network} shows the maximum hand position deviation when the number of robots in the formation is increased. To obtain such a figure, we start with a formation of $1$ circle and increase at each simulation the number of circles in the formation to $10$ circles. We recorded at each simulation the maximum hand position deviation for each circle and finally plot the maximum deviation on each circle across all the simulations. In accordance with our theoretical predictions, the figure shows that  disturbances are not amplified through the network. To further validate the results, we also report in Figure \ref{fig:hand_position_deviation} the hand position deviation of all robots when one robot in circle $1$ is affected by $d_i(t)$ in \eqref{equ: numerical_disturbance}. As expected, our protocol is able to reject the ramp component of the disturbance and, at the same time, prohibit the amplification of the residual component $w_i(t)$ in \eqref{equ: numerical_disturbance}.

\section{Conclusions}

We considered the problem of designing distributed control protocols for network systems affected by delays and disturbances. We proposed to leverage a multiplex architecture so that: (i) polynomial disturbances up to ramps are rejected; (ii) the amplification of  residual disturbances is prohibited. We then gave a sufficient condition on the control protocols to guarantee the fulfillment of these properties. The effectiveness of the result was illustrated, via simulations, on the problem of controlling the formation of unicycle robots.  Our future work includes extending the multiplex architecture and the results presented here to reject higher order polynomial disturbances.

%\bibliography{Reference_network}

\bibliographystyle{ifacconf}
\end{document}